\documentclass[preprint,showkeys,showpacs,preprintnumbers,amsmath,amssymb]{revtex4}

\usepackage{graphicx}  % Include figure files
\usepackage{dcolumn}   % Align table columns on decimal point
\usepackage{bm}        % bold math

\usepackage{amssymb, amsmath}
\usepackage{mathrsfs}			% required for \lap := \mathscr L

\newcommand{\mapping}{{\Phi}}		% transformation
 \newcommand{\ve}[1]{{\bf #1}}
 \newcommand{\vve}[1]{{\bf #1}}
 \newcommand{\vveg}[1]{{\boldsymbol {#1}}}	% Greek matrix

\newcommand{\matthree}[9]{\bracket{\begin{array}{ccc}
		#1	&#2	&#3	\\
		#4	&#5	&#6	\\
		#7	&#8	&#9
		\end{array}}}

\newcommand{\vecthree} [3] {\bracket{\begin{array}{c}
		#1 \\
		#2 \\
		#3
		\end{array}} }

\newcommand{\D}{\partial}

\newcommand{\dt}[1]{\frac {d #1} {d t}}
\newcommand{\delt}{{\Delta t}}

\newcommand{\Di}[1]{\frac {\D #1} {\D x_i}}

\newcommand{\DiDi}[1]{\frac {\D^2 #1} {\D x_i^2 }}

\newcommand{\excl}[1]{{\backslash \hspace{-0.3em} #1}}

\newcommand{\abs}[1]{\left|#1\right|}
\newcommand{\bracket}[1]{\left[#1\right]}

\newcommand{\parenth}[1]{\left(#1\right)}

\DeclareSymbolFont{AMSb}{U}{msb}{m}{n}
\DeclareMathSymbol{\R}{\mathbin}{AMSb}{"52}

\begin{document}

\title{Quantitative causality analysis with coarsely sampled time series}
        \author{X. San Liang}
        \date{\today}

\begin{abstract}
The information flow-based quantitative causality analysis has been widely 
applied in different disciplines because of its origin 
from first principles, its concise form, and its computational efficiency.
So far the algorithm for its estimation 
is based on differential dynamical systems, which, however, 
may make an issue for coarsely sampled time series.
Here, we show that for linear systems, this is fine at least
qualitatively; but for highly nonlinear systems, the bias increases
significantly as the sampling frequency is reduced.
This paper provides a partial solution to this problem, showing how
causality analysis is assured faithful with coarsely sampled series
when, of course, the statistics is sufficient. 
An explict and concise formula has been obtained, with only 
sample covariances involved.
It has been successfully applied to a system comprising of a pair 
of coupled R\"ossler oscillators. Particularly remarkable is the success
when the two oscillators are nearly synchronized.

\end{abstract}


\maketitle

\section{Introduction}
Causality analysis is an important problem in scientific research. Though
traditionally formulated as a statistical problem in data science, computer
science, among other disciplines, 
recently it has been found to be, within the framework of 
information flow/transfer,  ``a real notion in physics that 
can be derived {\it ab initio}''\cite{Liang2016}. A comprehensive study 
with generic systems has been fulfilled recently, with explicit formulas 
attained in closed form; see \cite{Liang2008} and \cite{Liang2016}. 
These formulas 
have been validated with benchmark systems such as baker transformation, 
H\'enon map, etc., and have been applied successfully 
to real world problems in the 
diverse disciplines such as 
global climate change 
(e.g., \cite{Stips2016}, \cite{Vannitsem2019}, \cite{Docquier2022}),
dynamic meteorology (e.g.,\cite{Liang2019}),
land-atmosphere interaction (e.g.,\cite{Hagan2018}),
data-driven prediction (e.g., \cite{Bai2018}, \cite{LiangXuRong2021}), 
near-wall turbulence (e.g., \cite{LiangLozano2016}),
neuroscience (e.g., \cite{Hristopulos2019}, \cite{Zhangtao2023}), 
financial analysis (e.g., \cite{Lu2022}, \cite{Lu2023}), 
quantum information (e.g, \cite{YiBose2022}), 
to name several.

For the purpose of this study, 
we first give a brief introduction of the theory within
the framework of a differential dynamical system.  (Also available 
for discrete-time mappings, refer to Liang (2016).)
Let
	\begin{eqnarray}	\label{eq:stoch_gov}
	\dt {\ve x} = \ve F(\ve x, t) + \vve B(\ve x, t) \dot {\ve w},
	\end{eqnarray}
be a $d$-dimensional continuous-time stochastic system 
for $\ve x = (x_1, ..., x_d)$ (we do not distinguish notations for random
and deterministic variables), where $\ve F = (F_1,..., F_d)$ 
may be arbitrary nonlinear differentiable functions of $\ve x$ and $t$,
${\ve w}$ is a vector of white noises, and $\vve B  = (b_{ij})$ 
is the matrix of perturbation amplitudes 
which may also be any differentiable functions of $\ve x$ and $t$.
Liang (2016)\cite{Liang2016} proves that
the rate of information flowing 
from $x_j$ to $x_i$ (in nats per unit time) is
	\begin{eqnarray}	\label{eq:Tji}
	T_{j\to i} 
        &=& -E \bracket{\frac1{\rho_i} 
		       \int_{\R^{d-2}} \Di{(F_i\rho_{\excl j})} 
				d\ve x_{\excl i \excl j}} + 
	     \frac 12 E \bracket{\frac1{\rho_i} 
	    \int_{\R^{d-2}} \DiDi {(g_{ii}\rho_{\excl j})} 
				d\ve x_{\excl i \excl j}}, \cr
	&=&
	- \int_{\R^d} \rho_{j|i} (x_j|x_i) \Di {(F_i\rho_{\excl j})} d\ve x
	  +	
	 \frac12 \int_{\R^d} \rho_{j|i} (x_j|x_i) 
		\DiDi {(g_{ii}\rho_\excl j)} d\ve x,
	\end{eqnarray}
	where $d\ve x_{\excl i \excl j}$ signifies 
	$dx_1 ... dx_{i-1} dx_{i+1} ... dx_{j-1} dx_{j+1}... dx_n$,
	$E$ stands for mathematical expectation, 
	$g_{ii} = \sum_{k=1}^n b_{ik} b_{ik}$, 
	$\rho_i = \rho_i(x_i)$ is the marginal probability density function
	(pdf) of $x_i$, $\rho_{j|i}$ is the pdf of $x_j$ conditioned on $x_i$,
	and $\rho_{\excl j} = \int_\R \rho(\ve x) dx_j$. 
The algorithm for the information flow-based causal inference is  
as follows: If $T_{j\to i} = 0$, then $x_j$ is not causal to $x_i$; 
otherwise it is causal, and the absolute value measures the magnitude 
of the causality from $x_j$ to $x_i$. This is guaranteed by a property
called ``principle of nil causality.''
Another property regards the invariance upon coordinate transformation,
indicating that the obtained information flow (IF) is an intrinsic property in
nature\cite{Liang2018}.
Also established by Liang (2016)\cite{Liang2016} is that, 
for a linear model, i.e., for
	$\ve F(\ve x, t) = \vve A \ve x,$
$\vve A = (a_{ij})$ and $\vve B = (b_{ij})$ are constant matrices 
in (\ref{eq:stoch_gov}), then
	\begin{eqnarray*}
	T_{j\to i} = a_{ij} \frac {\sigma_{ij}} {\sigma_{ii}}
	\end{eqnarray*}
where $\sigma_{ij}$ is the population covariance of $x_i$ and $x_j$.
By this, in the linear sense, causation implies correlation, but not vice
	versa.
In an explicit expression, this corollary fixes the debate on causation vs.
correlation ever since George Berkeley (1710)\cite{Berkeley1710}.

In the case with only $d$ time series $x_1, x_2,...,x_d$, 
the quantitative causality, i.e., the IF, between them can be estimated using 
maximum likelihood estimation (see \cite{Liang2014} and \cite{Liang2021}).
Under the assumption of a linear system with additive noises,
the maximum likelihood estimator (mle) of (\ref{eq:Tji}) for $T_{j\to i}$
is\cite{Liang2021}
	\begin{eqnarray}	\label{eq:T21_est}
	\hat T_{j\to i} = \frac 1 {\det\vve C} \cdot 
		       \sum_{\nu=1}^d \Delta_{j\nu} C_{\nu,di}
			\cdot \frac {C_{ij}} {C_{ii}},
	\end{eqnarray}
where $C_{ij}$ is the sample covariance between $x_i$ and $x_j$, 
	$\Delta_{ij}$ the cofactors of the matrix $\vve C=(C_{ij})$,
	and $C_{i,dj}$ the sample covariance between $x_i$ and 
	a series derived from $x_j$ using the Euler forward differencing
	scheme:
	$\dot x_{j,n} = (x_{j,n+k} - x_{j,n}) / (k\delt)$, with $k\ge1$ 
	some integer.
Eq.~(\ref{eq:T21_est}) is rather concise in form, involving
only the common statistics, i.e., sample covariances. 
The transparent formula makes causality analysis, which otherwise 
would be complicated, very easy and computationally efficient. 
Note, however, that Eq.~(\ref{eq:T21_est}) cannot replace (\ref{eq:Tji}); 
it is just the maximum likelihood estimator (mle) of the latter. 
Statistical significance tests can be performed for the estimators. This is done
with the aid of a Fisher information matrix. 
See Liang (2014)\cite{Liang2014} and Liang (2021)\cite{Liang2021} for details.

%As said above, this concise formula (\ref{eq:T21_est}) has 
%shown its success in a wide variety of fields. 
%Albeit the estimator of a linear version of (\ref{eq:Tji}), 
%it has demonstrated its performance for highly nonlinear sytems.

Originally the formalism is established in the light of a differential
system; in other words, it is with infinitesimal time increments. (The
formalism with discrete mappings has also been established by Liang
(2016)\cite{Liang2016}, but still there has no estimation with it.) 
One would naturally ask a question about the applicability in the case of
coarsely sampled time series. Indeed, it is not unusual that the given
series may be coarsely sampled because of the limited observations.
% anyhow, not all the measurements are like 
% those with the EEG series in neuroscience.
As will be seen in the following section this may make a problem for 
nonlinear systems if the sample interval is large. 
This paper henceforth attempts to address
this issue in the original linear framework. 
In the following we first check the applicability of (\ref{eq:T21_est})
for series from a linear system and a highly nonlinear system 
(section \ref{sect:TheIssue}), 
with a variety of sampling intervals. A new approach is presented in 
section~\ref{sect:method}, which is then utilized
to redo the causal inferences in section~\ref{sect:TheIssue}. 
Some remaining issues are discussed in section~\ref{sect:discussion}.

\section{The issue with coarsely sampled series}  \label{sect:TheIssue}
%In this section we use two examples to test the applicability of
%	Eq.~\ref{eq:T21_est}
%in the case of large sampling interval. The first is a well studied linear
%system whose known information flow rates have been found in a half
%analytical way. The second is a challenging one, with high nonlinearity and
%almost synchronized series.

\subsection{Time series from linear systems}
\begin{figure}[h]
	\begin{center}
	\includegraphics[width=1\textwidth]{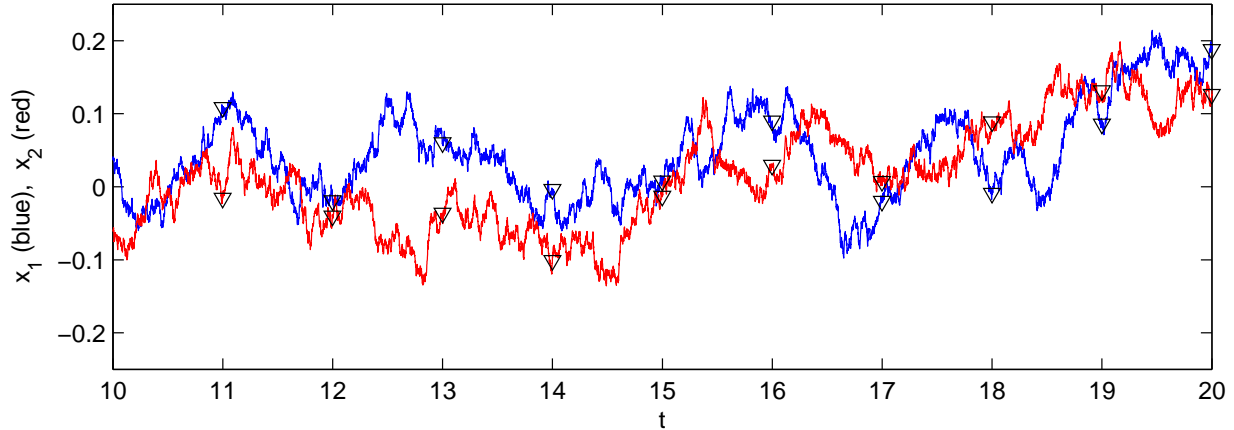}
	\caption{A segment of the sample path generated by 
	Eq.~(\ref{eq:liang2014_gov}), with a time step $\Delta t = 0.001$. 
	The time spans from 0 through 100. As we are only interested in the
	limit IF as $t\to\infty$, the first 5000 steps, i.e., those for 
	$t\le5$, are discarded.
	\protect{\label{fig:sample1000}} }
	\end{center}
	\end{figure}

We first test the applicability of (\ref{eq:T21_est}), as the sampling
interval increases, with a well-studied linear system whose IF rates have
been found half-analytically. This is the validation example in 
\cite{Liang2014}:
	\begin{subequations}	\label{eq:liang2014_gov}
 	\begin{eqnarray}
	\dt {x_1} &=&  -x_1 + 0.5 x_2 + 0.1 \dot w_1	\\
	\dt {x_2} &=& -x_2 + 0.1 \dot w_2
	\end{eqnarray}
	\end{subequations}
where $\dot w_i$, $i=1,2$ are independent white noises.   
It has been shown that, the rates of 
information flow per unit time, $T_{2\to1} \to 0.11$ as $t\to\infty$,
and $T_{1\to2} = 0$ for all $t$, reflecting accurately the one-way
causality from $x_2$ to $x_1$. Now, using the same sample path as that in
Liang (2014),\footnote
    {Note, because of the pseudorandom number generator, the generated
sample path using the normal differencing scheme may not be satisfactory. 
To see whether the obtained sample path is correct, one may
check the resulting covariances, which can be rather
accurately obtained by solving a deterministic ODE. Here the data 
for the generated sample path can be downloaded from 
http://www.ncoads.org/article/show/68.aspx under the item PRE\_2014.dat.
    } 
we re-sample the series with low frequencies to obtain new series. 
Shown in Figure~\ref{fig:sample1000} is part of the sample path,
with triangles marking the sampling points.

The computed IFs for different sampling intervals are listed below:
%
%%\begin{table}[h]
%\begin{center}
%%\caption{  \protect{\label{tab:}}}
%\begin{tabular}{ccc}
%\hline
%\hline
%Samp. interval	&  [T21, T12] old scheme      &	[T21,T12] new scheme \\
%\hline
%2000		& [0.016$\pm$0.030, 0.032$\pm$0.033]	& [0.126,  0.210] \\
%1000		& [0.061$\pm$0.041, -0.002$\pm$0.043]	& [0.166,  0.005] \\
%500		& [0.055$\pm$0.034, -0.015$\pm$0.034]	& [0.098, -0.015] \\
%300		& [0.055$\pm$0.036, -0.008$\pm$0.038]	& [0.082, -0.002] \\
%100		& [0.090$\pm$0.048, -0.011$\pm$0.053]	& [0.106, -0.002] \\
%50		& [0.099$\pm$0.050, -0.015$\pm$0.054]	& [0.109,  -0.007]\\
%10		& [0.113$\pm$0.051, -0.001$\pm$0.055]	& [0.118, 0.008]  \\
%1	        & [0.110$\pm$0.051, -0,002$\pm$0,056] & [0.114,  0.007] \\
%\hline
%\end{tabular}
%\end{center}
%%\end{table}
%
\begin{center}
\begin{tabular}{ccccccc}
\hline
\hline
S.I. (\# of pts)        & 1     & 10    & 50   &100   & 300   & 500   \\
$\hat T_{2\to1}$ &0.110$\pm$0.051 &0.113$\pm$0.051  &0.099$\pm$0.050 
	    &0.090$\pm$0.048 &0.055$\pm$0.036  &0.055$\pm$0.034   \\
$\hat T_{1\to2}$ &-0.002$\pm$0.056 &-0.001$\pm$0.055  &-0.015$\pm$0.054 
	    &-0.011$\pm$0.053 &-0.008$\pm$0.038  &-0.015$\pm$0.034 \\
\hline
\end{tabular}
\end{center}
Also computed are the confidence intervals at a level of 90\% (at a significance
level of 0.1). First, the estimators $\hat T_{2\to1}$ for all the SIs here
are significantly distinct from zero, while those the other way
around, $\hat T_{1\to 2}$, are not significant at a level of 90\%.
So the causality in a qualitative sense has been faithfully recovered even
with very low sampling frequencies (large SI). (In fact, even with SI=1000
the result is still correct; we do not consider cases beyond SI=500 since 
the sample size is too small for SI>500, resulting in insufficient statistics.)

Since this example actually has a half-analytical solution 
	($T_{2\to1}\approx0.11$, $T_{1\to2}=0$), 
we have more to say about the computed results.
Generally, the result of $\hat T_{1\to2}$ looks satisfactory.
For $\hat T_{2\to1}$, it is rather accurate for SI $\le$ 100. 
Beyond 100, it is not accurate any more. 

\subsection{Time series from synchronized chaotic oscillators}

\begin{figure}[h]
	\begin{center}
	\includegraphics[width=0.75\textwidth]{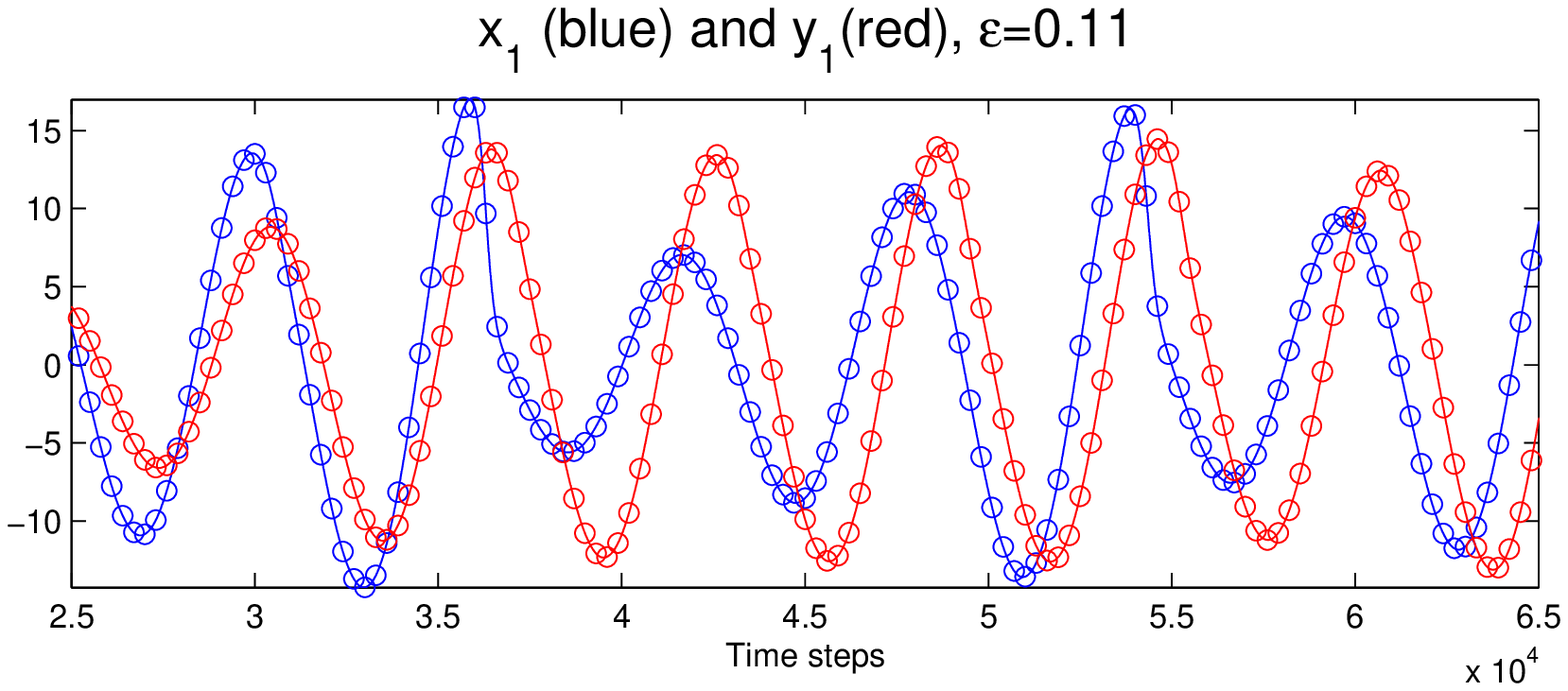}
	\includegraphics[width=0.75\textwidth]{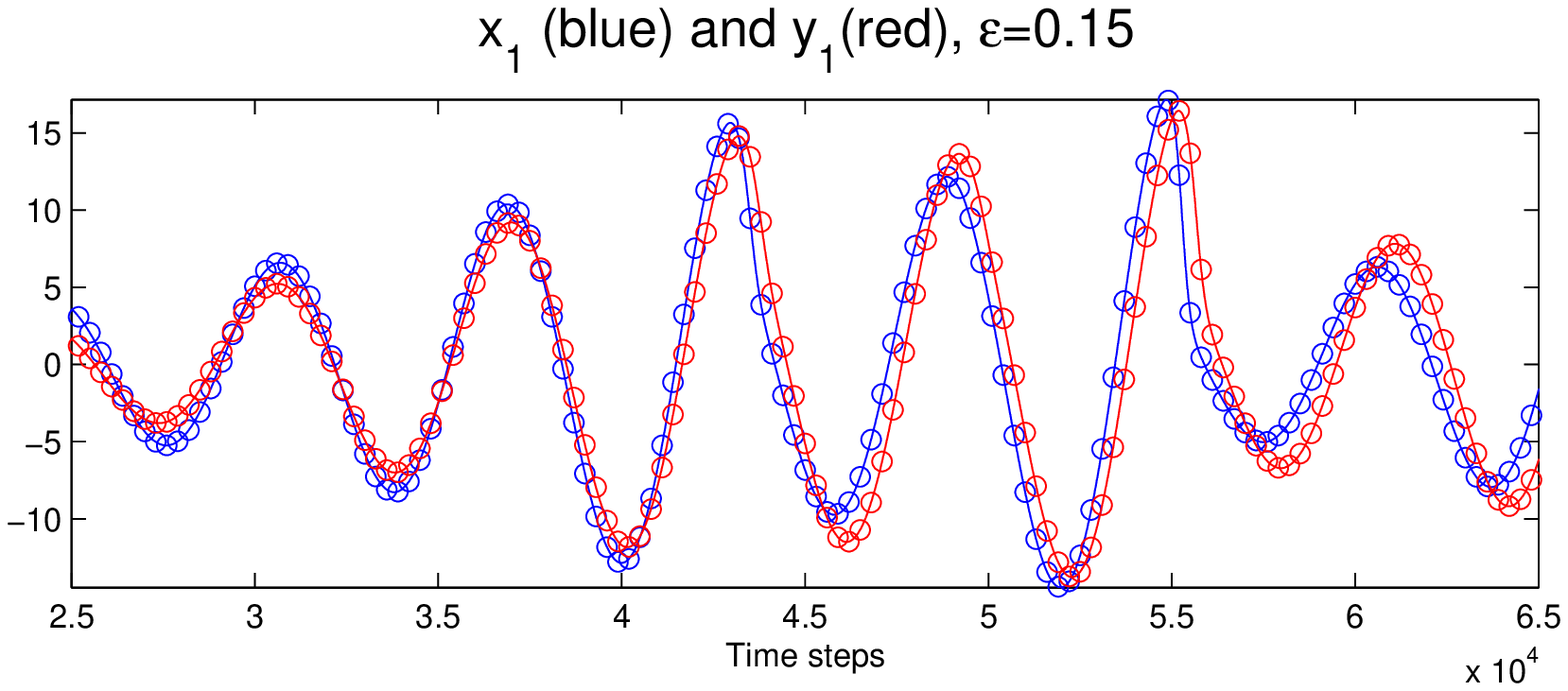}
	\caption{Part of the time series of $x_1$ and $y_1$ of the coupled
	R\"ossler systems for $\varepsilon=0.11$ (top) and
	$\varepsilon=0.15$ (bottom). 
	The circles indicate
	the sampling points (every 300 steps here, corresponding to
	18 points in each period). 
	The two oscillators become synchronized as $\varepsilon\ge 0.15$. 
	\protect{\label{fig:xy_SI500}} }
	\end{center}
	\end{figure}

The following example is from the synchronization problem as examined by
Palus et al. (2018). The system is composed of two R\"ossler oscillators,
$\ve x = (x_1, x_2, x_3)$ and $\ve y = (y_1, y_2, y_3)$, where 
	\begin{subequations}
	\begin{eqnarray}
	&&\dt {x_1} = -\omega_1 x_2 - x_3,	\\
	&&\dt {x_2} = \omega_1 x_1 + 0.15 x_2,	\\
	&&\dt {x_3} = 0.2 + x_3 (x_1 - 10),
	\end{eqnarray}
	\end{subequations}
is the master system, and 
	\begin{subequations}
	\begin{eqnarray}
	&&\dt {y_1} = -\omega_2 y_2 - y_3 + \varepsilon (x_1 - y_1),	\\
	&&\dt {y_2} = \omega_2 y_1 + 0.15 y_2,	\\
	&&\dt {y_3} = 0.2 + y_3 (y_1 - 10),
	\end{eqnarray}
	\end{subequations}
is the driven one. Following Palus et al. (2018)\cite{Palus2018}, 
choose $\omega_1 = 1.015$
and $\omega_2 = 0.985$. Using the Runge-Kutta scheme and choosing 
a time step $\Delta t= 0.001$, the coupled 6-dimensional system can 
be solved rather accurately with different $\varepsilon$.
Figure~\ref{fig:xy_SI500} plots the solutions of $x_1$ and $y_1$ 
when the coupling strength $\varepsilon=0.11$ (upper panel)
and $\varepsilon=0.15$ (lower panel). 
As shown in the latter case, the two subsystems become synchronized 
if $\varepsilon\ge0.15$.
Again, we choose to study the problem for $t\in[0,100]$ ($10^5$ time steps
in total).

\begin{figure}[h]
	\begin{center}
	\includegraphics[width=0.49\textwidth]{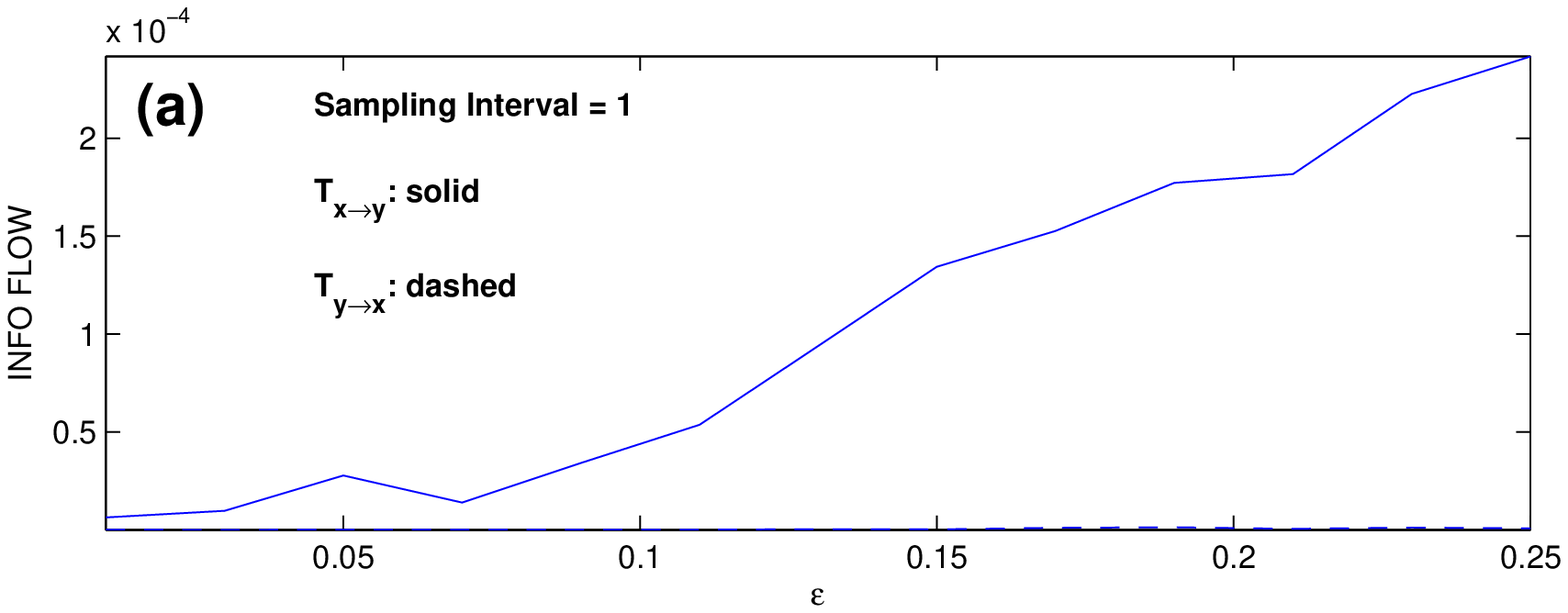}
	\includegraphics[width=0.49\textwidth]{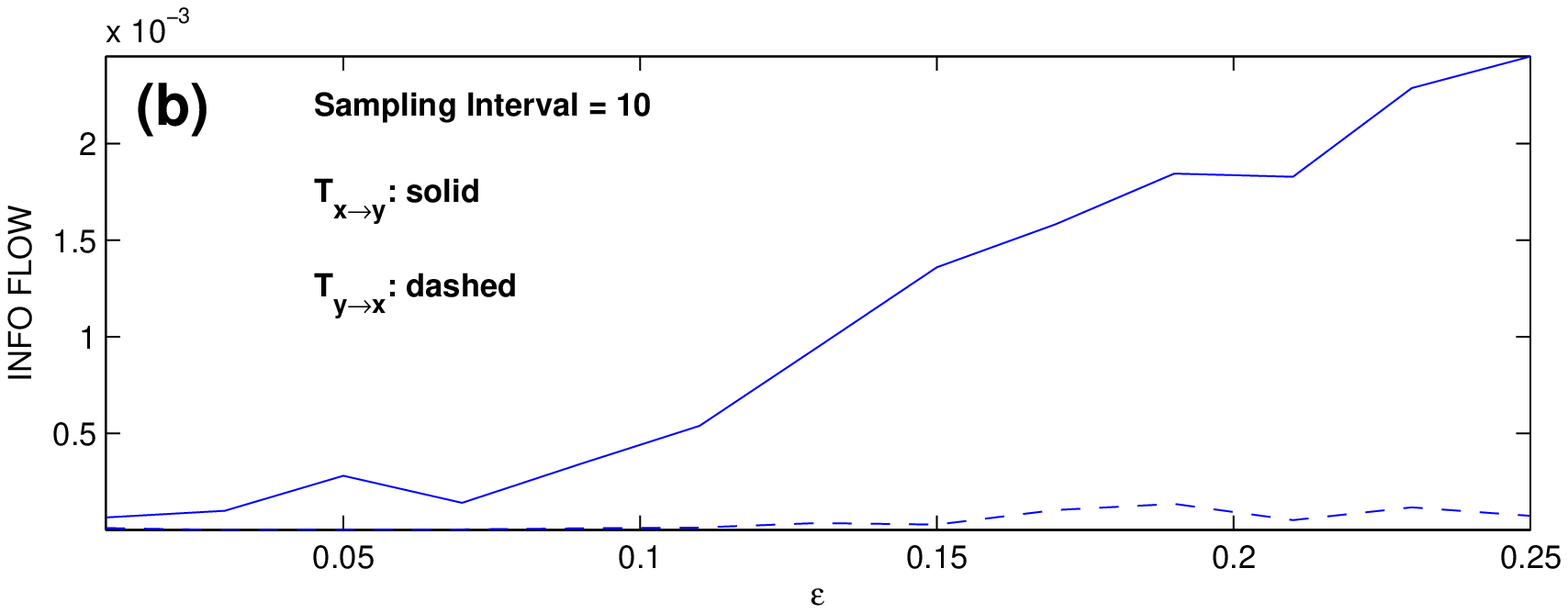}
	\includegraphics[width=0.49\textwidth]{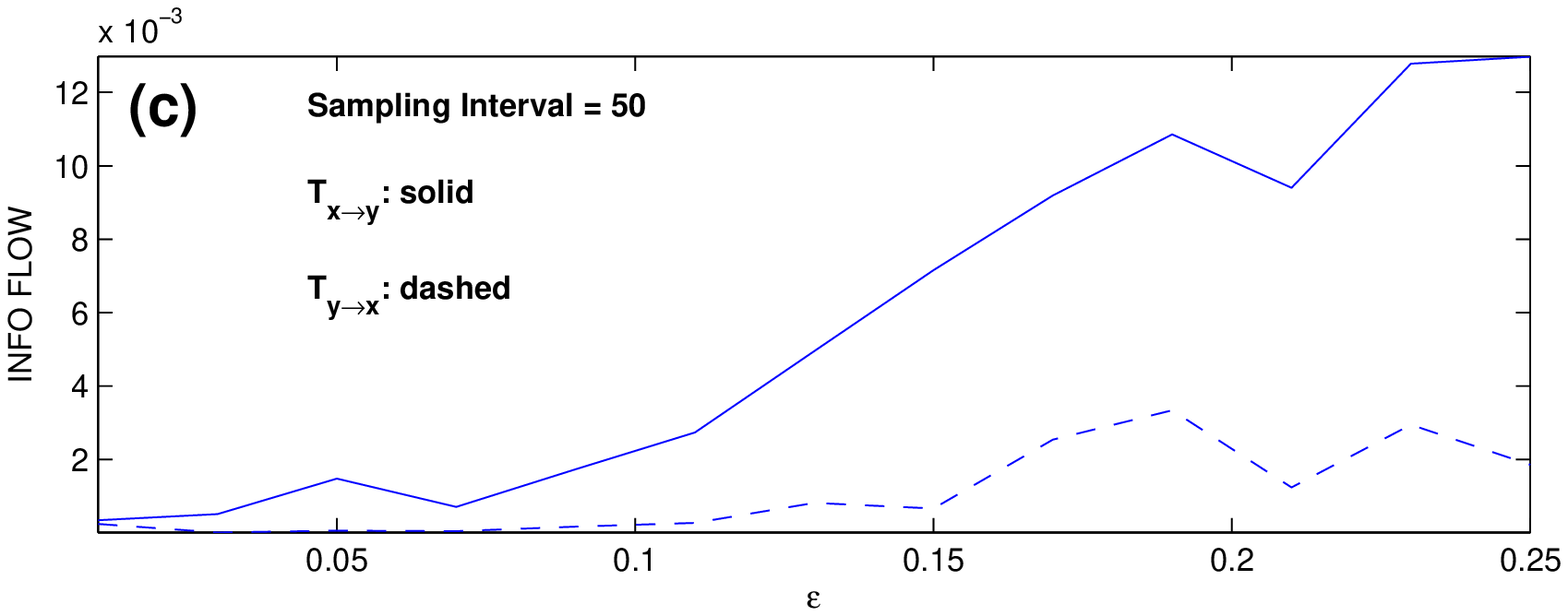}
	\includegraphics[width=0.49\textwidth]{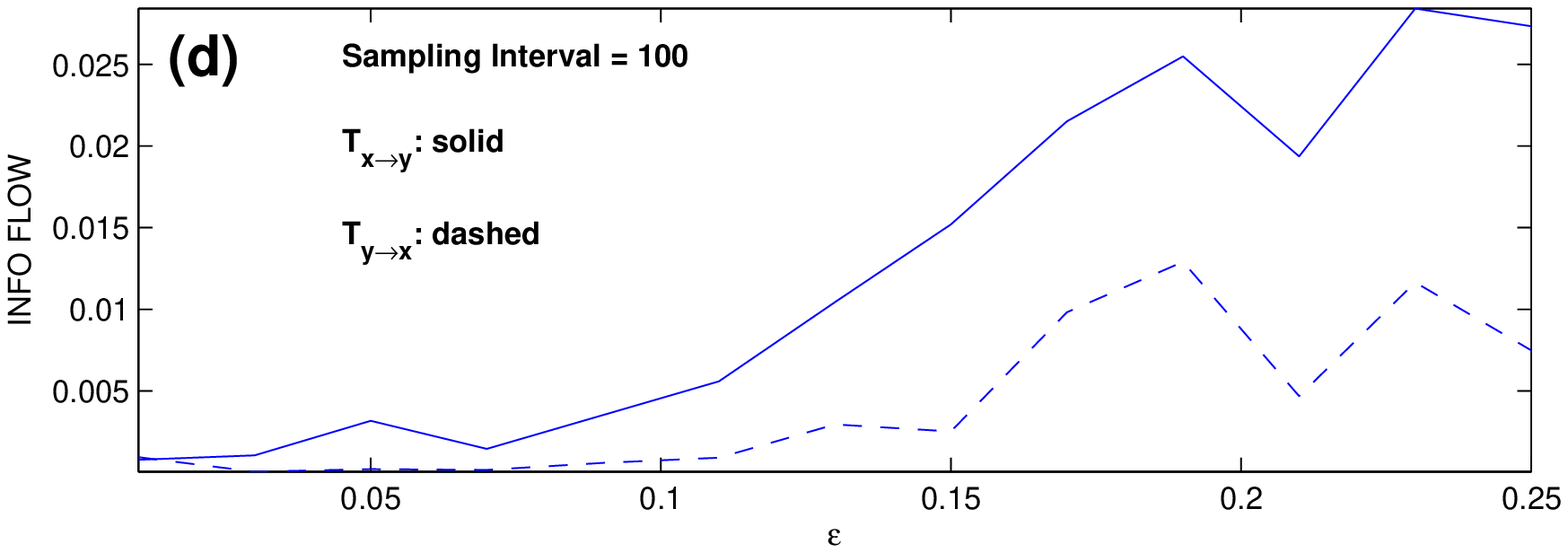}
	\includegraphics[width=0.49\textwidth]{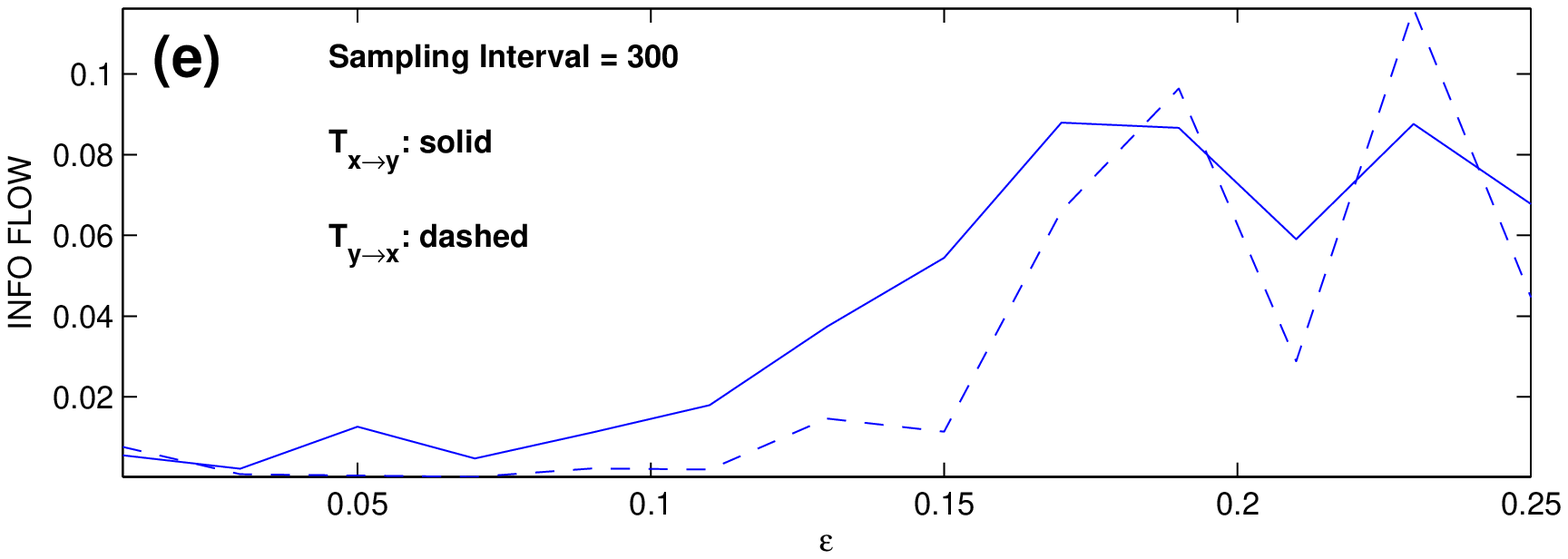}
	\includegraphics[width=0.49\textwidth]{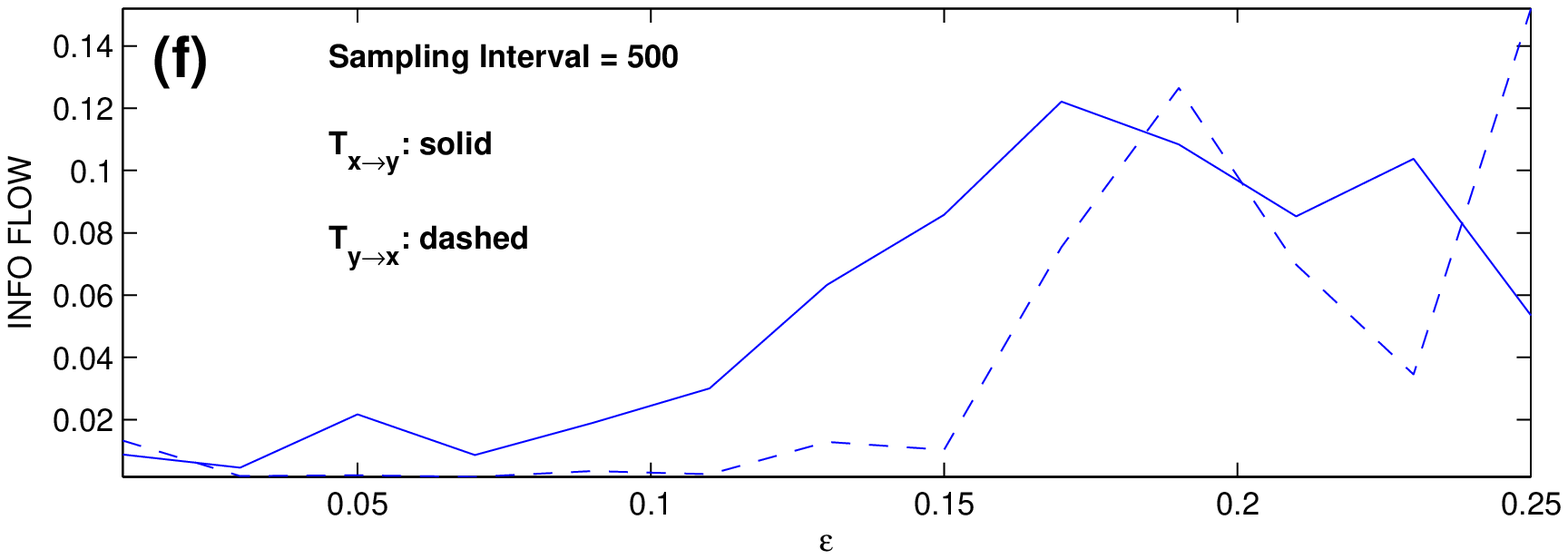}
	\caption{Absolute information flow rates 
	$\abs{\hat T_{x\to y}}$ and $\abs{\hat T_{y\to x}}$ (dashed) between the two R\"ossler
	oscillators as functions of the coupling
	coefficient $\varepsilon$ with different sampling intervals. 
	Units are in nats per unit time. By the preset causality, the
	dashed line should coincide with the abscissa.
	\protect{\label{fig:IF_rossler}} }
	\end{center}
	\end{figure}

For each $\varepsilon$ we generate six time series of $10^5$ steps,
and evaluate the IFs according to Eq.~(\ref{eq:T21_est}) ($k=1$ is chosen). 
The IFs as functions of $\varepsilon$ are then obtained, and
plotted in Fig.~\ref{fig:IF_rossler}a, which accurately tells that 
the master is $\ve x$, and $\ve y$ is the slave. An appealing observation
is that this causality inference even works when the two oscillators 
are nearly synchronized as $\varepsilon>0.15$, demonstrating the power
of this rigorously formulated causality analysis.

We subsample the series every SI steps,  SI=10, 50, 100, 300, 500, 
and redo the computation using the same scheme. The resulting IF
rates are shown in Figs.~\ref{fig:IF_rossler}b, c, d, e,
f, respectively. By the preset causality, the dashed line should be the
zero-line. Clearly, the causal inference works well for SI$\le10$.
The computed IF becomes biased for SI$\ge50$, and the bias grows
significantly as SI increases. 
If we focus on $\varepsilon<=0.15$, i.e., when the systems
are not synchronized (see Palus et al., 2018), the causal inference 
still functions fine for SI$\le$50. 
If the synchronized cases are taken into account ($\varepsilon>0.15$), 
then the inferences in the cases for $50\le {\rm SI} \le 100$ are much
biased, and those for SI exceeding 300, a case corresponding to an
approximate sampling frequency of 20 per period, are not correct any more.

\section{Approaching to a partial solution}	\label{sect:method}

As shown above, if the sampling frequency of the time series is low, 
the resulting linear IF for nonlinear series may be biased.
Indeed, in the case with high nonlinearity, the linear assumption 
is always easy to be blamed. 
While theoretically it is not a problem (causality is guaranteed as proved
in a theorem), we agree that, before a fully nonlinear algorithm is
developed, this will be a continuing issue.
What we want to show here is, how much room there is for improvement.
So far, the algorithm documented in \cite{Liang2014}, and later in 
\cite{Liang2021}, is based on the Bernstein-Euler differencing scheme, 
which is, of course, very rudimentary due to the first order differencing. 
If a time series is coarsely sampled, the error could be large.

A theorem as established by Liang (2008)\cite{Liang2008} 
reads that, if the 
noise is additive in Eq.~(\ref{eq:stoch_gov}), i.e., if $\vve B$ is a
constant matrix, then the noise itself does not appear in 
the formula of $T_{j\to i}$.
So under the additive noise assumption, we can estimate the 
IF within the framework of a deterministic system.
In this case, note that the linear equation set actually can
be solved for an interval $[t, t+\Delta t]$, no matter how large 
$\Delta t$ is. This gives a hint to the solution of the low sampling
frequency problem.

Consider 
	\begin{eqnarray}
	\dt {\ve x} = \ve f + \vve A \ve x,
	\end{eqnarray}
where $\vve A = (a_{ij})$ is a $d\times d$ matrix.
Let us assume that $\ve f = \ve 0$, since the time series can always be
pre-treated by removing the linear trend, and it has been proved that this
removal does not alter the IF rates. In this case, on the interval
$[t,t+\Delta t]$, we actually have a mapping $\mapping: \R^d \to \R^d$
that takes the state $\ve x(t)$ to the state $\ve x(t+\Delta t)$ 
at $t+\Delta t$, with the propagating operator:
	\begin{eqnarray}
	\Phi = e^{\vve A\Delta t} 
	     = e^{\matthree {a_{11}} \hdots {a_{1d}} 
		      	    \vdots   \ddots \vdots
			    {a_{d1}} \hdots {a_{dd}} \Delta t}
	    \equiv \matthree {\alpha_{11}} \hdots {\alpha_{1d}} 
			      \vdots       \ddots  \vdots
			     {\alpha_{d1}} \hdots {\alpha_{dd}}.
	\end{eqnarray}
It is not easy to estimate $a_{ij}$, but it is easy to estimate
$\alpha_{ij}$ instead, by observing the relation
	\begin{eqnarray}
	&&   \matthree {\alpha_{11}} \hdots {\alpha_{1d}} 
		      \vdots       \ddots  \vdots
		     {\alpha_{d1}} \hdots {\alpha_{dd}} 
		\ve x(n) = \ve x(n+1), 	\qquad\qquad
	     n = 0, 1, 2, ..., N.  
	\end{eqnarray}
This written in a matrix form is 
	\begin{eqnarray*}
	\matthree {x_1(0)} \hdots  {x_d(0)}
		  \vdots   \ddots  \vdots
		  {x_1(N-1)} \hdots {x_d(N-1)}
	\vecthree {\alpha_{i1}}  {\vdots}  {\alpha_{id}}
	=
	\vecthree {x_i(1)} \vdots {x_i(N)}
	\end{eqnarray*}
for $i=1,..,d$. 
Averaging all the rows of the algebraic equation set, and subtracting the
mean from each row, we get
	\begin{eqnarray*}
	\matthree {x_1(0)- \bar x_1} \hdots  {x_d(0)- \bar x_d}
		  \vdots   \ddots  \vdots
		  {x_1(N-1)- \bar x_1} \hdots {x_d(N-1)- \bar x_d}
	\vecthree {\alpha_{i1}}  {\vdots}  {\alpha_{id}}
	=
	\vecthree {x_i(1)- \bar x_{i+}} \vdots {x_i(N)- \bar x_{i+}},
	\end{eqnarray*}
where $\bar x_i = \frac 1 N \sum_{n=0}^{N-1} x_i(n)$,
      $\bar x_{i+} = \frac 1 N \sum_{n=1}^N x_i(n)$,
i.e., the series $\{x_{i+}(n)\}$ is the series $\{x_i(n)\}$ 
advanced by one step.
Let $i$ run through $\{1,2,...,d\}$. We have the following $d$
overdetermined equation sets:
	\begin{eqnarray}	\label{eq:full_eqsets}
	\matthree {x_1(0)- \bar x_1} \hdots  {x_d(0)- \bar x_d}
		  \vdots   \ddots  \vdots
		  {x_1(N-1)- \bar x_1} \hdots {x_d(N-1)- \bar x_d}
	\matthree {\alpha_{11}}  \hdots  {\alpha_{d1}}
		  \vdots	 \ddots	   \vdots
	          {\alpha_{1d}}  {\vdots}  {\alpha_{dd}}
	=
	\matthree {x_1(1)- \bar x_{1+}} \vdots {x_d(1)- \bar x_{d+}}
		  \vdots	 \ddots	   \vdots
	          {x_1(N)- \bar x_{1+}} \vdots {x_d(N)- \bar x_{d+}}.
	\end{eqnarray}
Denote by $\vveg {\Lambda}$ the matrix $(\alpha_{ij})$, then the 
matrix of unknowns in the above equation sets is $\vveg {\Lambda}^T$. 
Left multiplication by
	\begin{eqnarray*}
	{\matthree {x_1(0)- \bar x_1} \hdots  {x_d(0)- \bar x_d}
		  \vdots   \ddots  \vdots
		  {x_1(N-1)- \bar x_1} \hdots {x_d(N-1)- \bar x_d}}^T
	\end{eqnarray*}
on both sides yields $d$ $d\times d$ equation sets:
	\begin{eqnarray}	\label{eq:leastsq}
	\vve C \vveg \Lambda^T = \vve {\tilde C},
	\end{eqnarray}
where $\vve C = (C_{ij})$ is the sample covariance matrix of $\ve x$, and
$\vve {\tilde C} = (C_{i,j+})$, and $C_{i,j+}$ is the sample covariance 
between $x_i$ and $x_{j+}$, i.e., $x_j$ advanced by one time step.
The least square solutions of the overdetermined sets (\ref{eq:full_eqsets}) 
are the solutions of (\ref{eq:leastsq}):
	\begin{eqnarray*}
	\vveg\Lambda^T = \vve C^{-1} \vve {\tilde C},
	\end{eqnarray*}
and hence
	\begin{eqnarray*}
	\vveg\Lambda = (\vve C^{-1} \vve {\tilde C})^T 
		     = \vve {\tilde C}^T \vve C^{-1}.
	\end{eqnarray*}
The estimator of $\vve A$ is, therefore,
	\begin{eqnarray}
	\vve {\hat A} = \frac 1 {\Delta t} 
		 \log \parenth{\vve {\tilde C}^T \vve C^{-1}}.
	\end{eqnarray}
(Caution should be used in case of singularity. The irrelevant imaginary
part also should be discarded.)

Once getting $A$, hence the coefficients $(a_{ij})$, we substitute 
$a_{ij}$ for the whole part 
	\begin{eqnarray*}
	\frac 1 {\det\vve C} \sum_{k=1}^d \Delta_{jk} C_{k,di}
	\end{eqnarray*}
in Eq.~(\ref{eq:T21_est}), i.e., multiply $a_{ij}$ by
$C_{ij}/C_{ii}$ to arrive at the desideratum, $\hat T_{j\to i}$. 
If we denote by $[\vve A]_{ij}$ the extraction of the $(i,j)^{th}$ entry of
the matrix $\vve A$, this is
	\begin{eqnarray}	\label{eq:Tji_est_new}
	\hat T_{j\to i} = 
	\frac 1 {\Delta t} \bracket{\log(\tilde {\vve C}^T \vve C^{-1})}_{ij}
		\cdot \frac {C_{ij}} {C_{ii}}.
	\end{eqnarray}
(Note here log is the matrix logarithm. In matlab, the function is logm.)

\section{The coarsely sampled series problems revisited}
As demonstrated above, for the series generated from linear systems, the
estimation of the IF is fine qualitatively. 
We here, nevertheless, want to see
how the new scheme may have the results improved. Shown below is a
recalculation of the estimates. Since this case has a rather accurate
result ($\approx 0.11$ nats per unit time), we can see that the
result is accurate enough for all the SIs here.

\begin{center}
\begin{tabular}{ccccccc}
\hline
\hline
Sampling Interval & 1     & 10    & 50   &100   & 300   & 500   \\
$\hat T_{2\to1}$ &0.114\  &0.118\  &0.109\  &0.106\ &0.082\  &0.098 \\
$\hat T_{1\to2}$ &0.007 &0.008 &$-0.007$ &$-0.002$ &$-0.002$ &$-0.015$ \\
\hline
\end{tabular}
\end{center}

\begin{figure}[h]
	\begin{center}
	\includegraphics[width=0.49\textwidth]{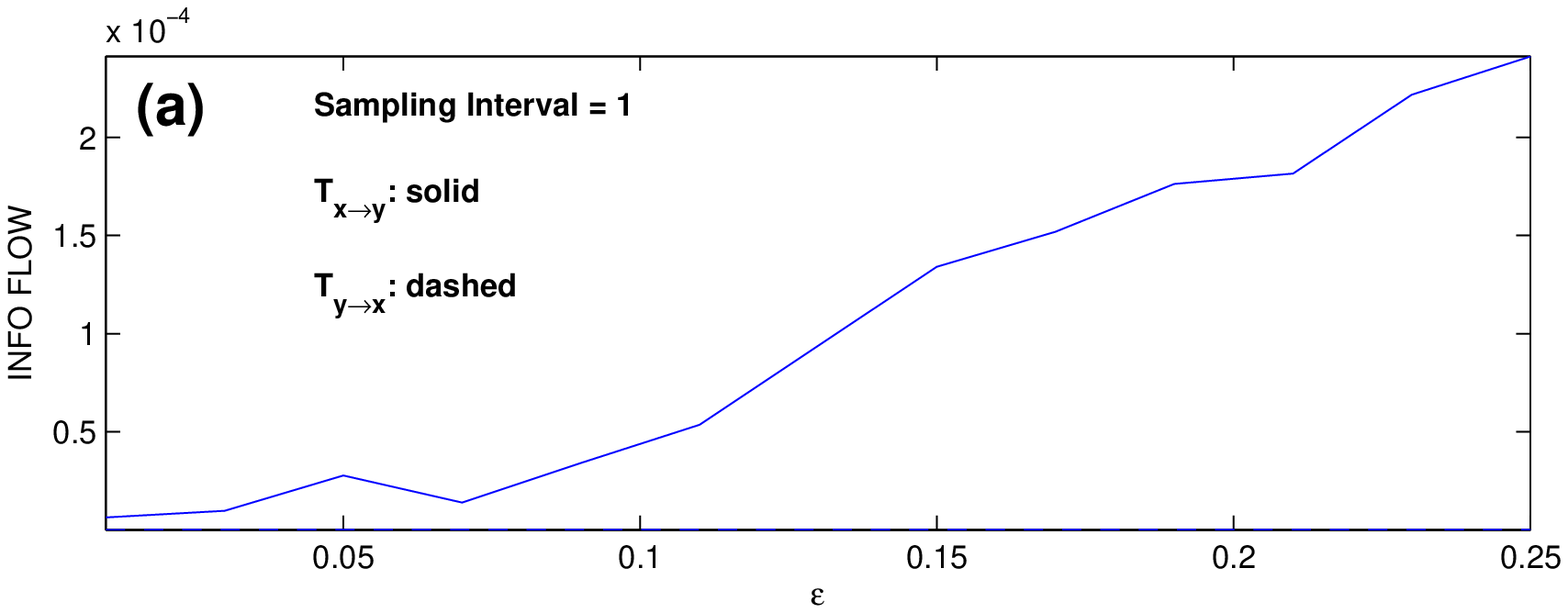}
	\includegraphics[width=0.49\textwidth]{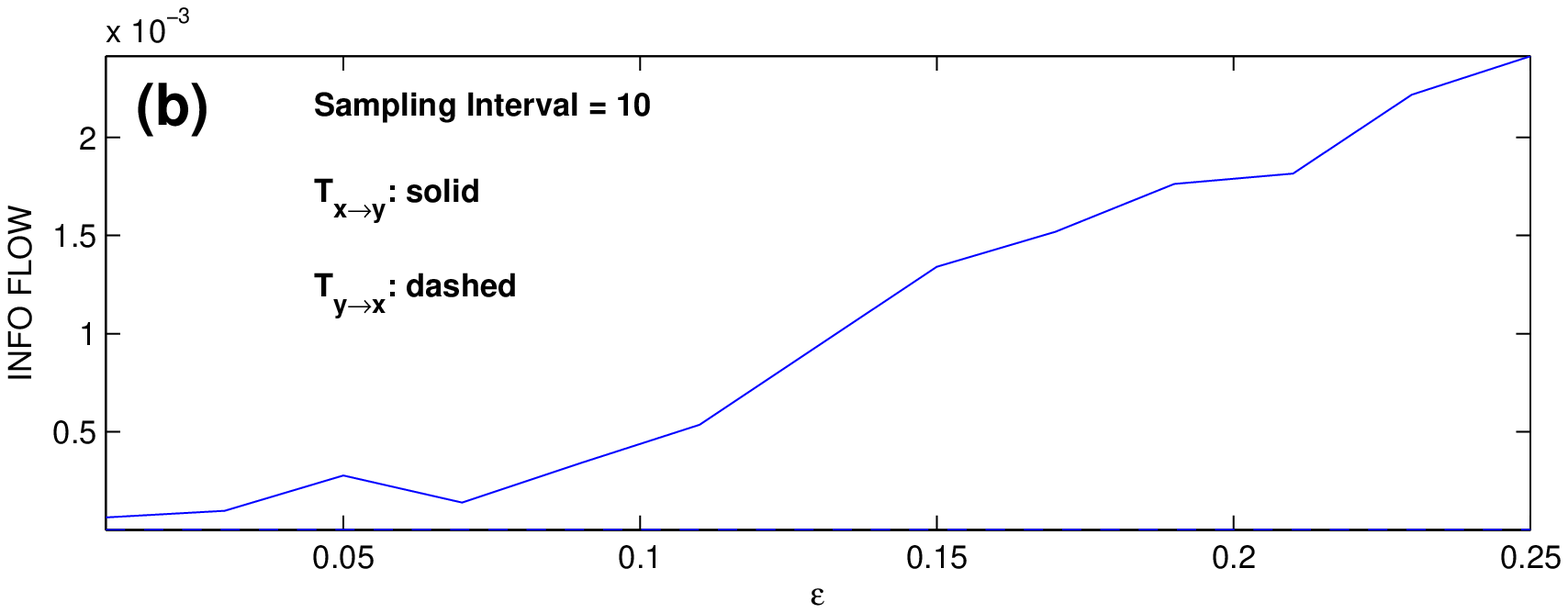}
	\includegraphics[width=0.49\textwidth]{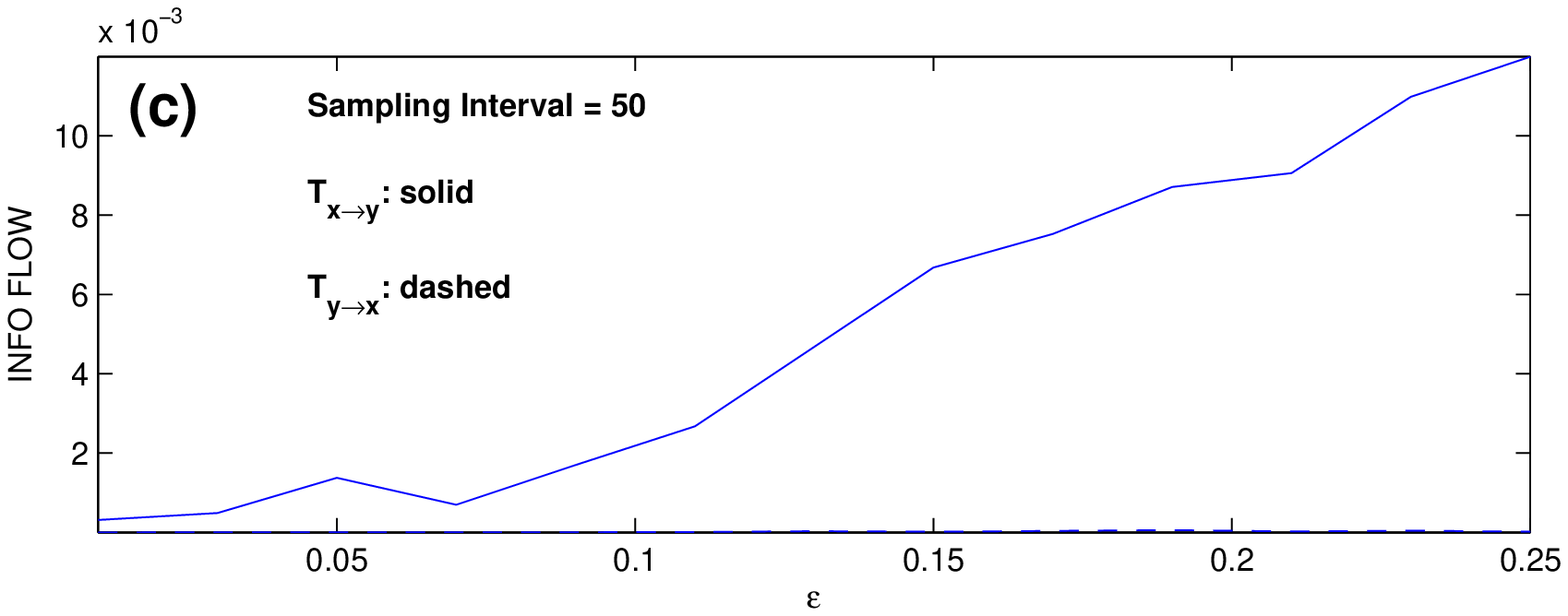}
	\includegraphics[width=0.49\textwidth]{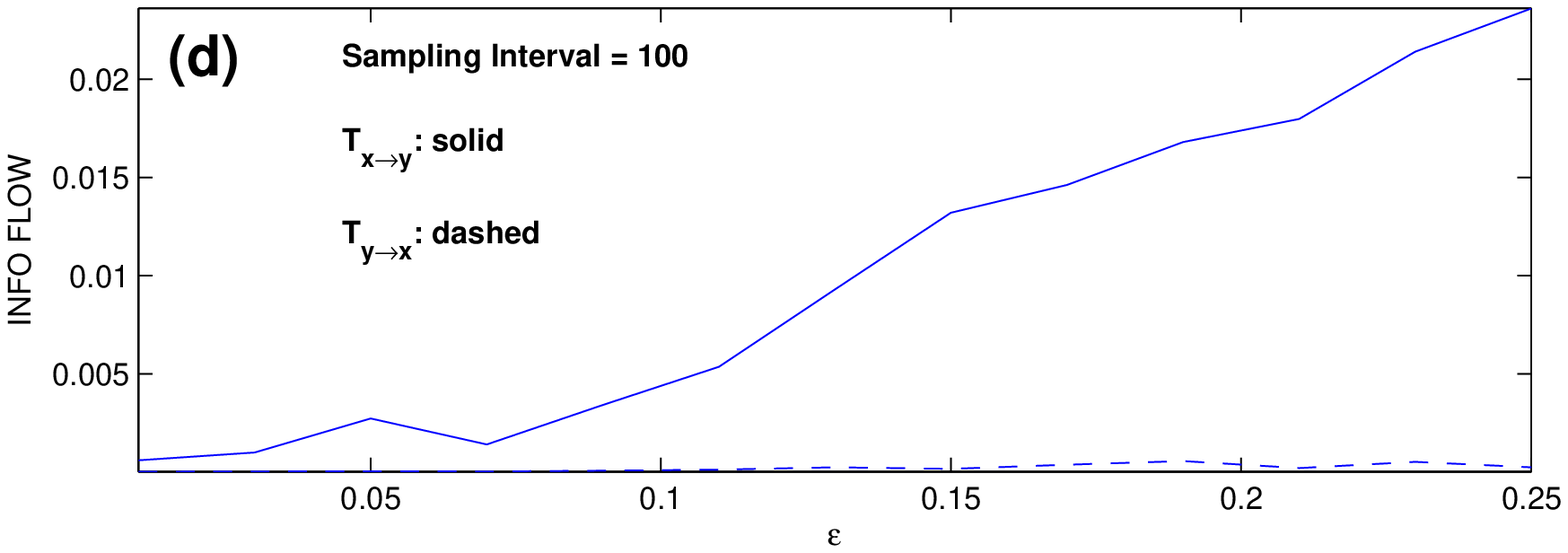}
	\includegraphics[width=0.49\textwidth]{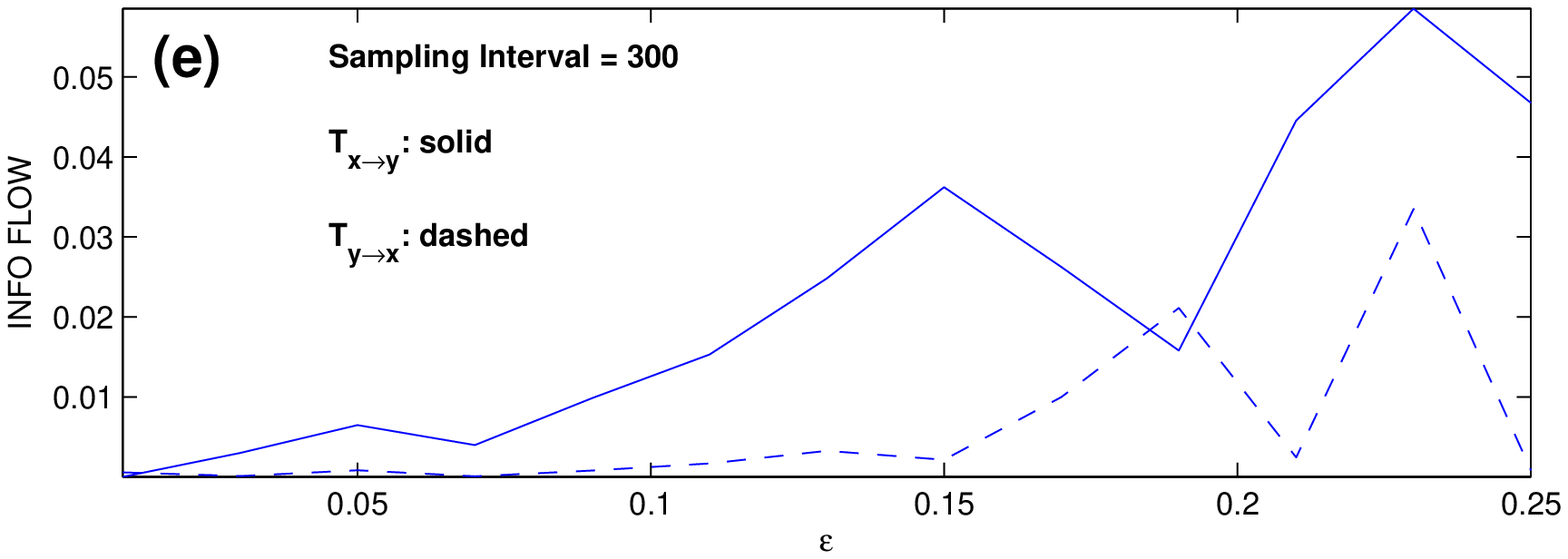}
	\includegraphics[width=0.49\textwidth]{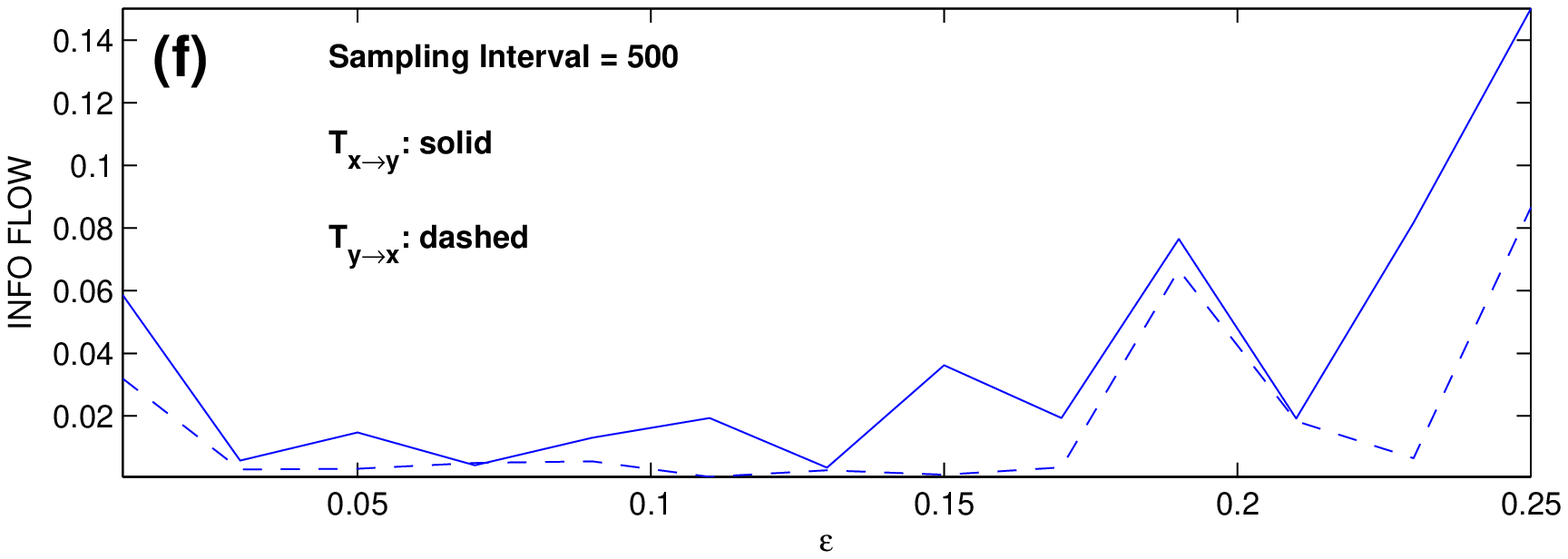}
	\caption{As Fig.~\ref{fig:IF_rossler}, but the information flow
	rates are computed with the new scheme. 
	\protect{\label{fig:new_IF_rossler}} }
	\end{center}
	\end{figure}

The new scheme for the estimation is particularly for the nonlinear case. For the
pair of R\"ossler oscillators, the computed results are plotted in
Fig.~\ref{fig:new_IF_rossler}. Compared to Fig.~\ref{fig:IF_rossler},
now the performance has been much improved. 
For the cases with SI$\le100$ (Figs.~\ref{fig:new_IF_rossler}a-d), 
the results are rather accurate for all
the coupling strengths $\varepsilon$ considered (both synchronized and 
non-synchronized). For the case SI=300, which corresponds to a sampling
frequency of 20 points per period, the one-way causality is accurately
recovered for the nonsynchronized cases ($\varepsilon\le0.15$). But beyond
that $\varepsilon>0.15$, the inference fails. Particularly, when SI=500
(Fig.~\ref{fig:new_IF_rossler}f),
the result is even worse that its counterpart with the traditional scheme
as plotted in Fig.~\ref{fig:IF_rossler}f. This, of course, may be due to
the resulting small sample size, which causes singularity to the 
matrix logarithm. 

\section{Discussion}	\label{sect:discussion}
The maximum likelihood estimator of the information flow (IF),
Eq.~(\ref{eq:T21_est}), provides a very easy way to causal inference.
Theoretically it is based on a linear assumption, but practically it has
shown tremendous success with series generated from highly nonlinear
systems; anyway, linearization piecewise in time proves to be an efficient
asymptote to an otherwise nonlinear system. In reality, series may be 
coarsely sampled; the time resolution may be low. An issue thus arises,
as this formalism is theoretically on the basis of infinitesimal time
increments. In this case, as we have shown, it still works for linear
systems in a qualitative sense; 
but for a highly nonlinear system composed of two R\"ossler
oscillators, the bias becomes more and more significant as the sampling 
frequency is reduced. 

A new scheme has been proposed to address this problem and provide a
partial solution. Due to the nice property of IF, as proved
in \cite{Liang2008}, that additive noises do not alter the IF flow in form, 
it is reasonable to directly estimate the IF without paying
attention to the stochasticity. Instead of estimating through the
differential equations using the Euler-Bernstein differencing, 
we choose to consider the integral form on the finite time interval, i.e.,
to estimate the Lie group members. 
In doing this, the original formula (\ref{eq:T21_est}), which is rewritten
here for easy reference, 
	\begin{eqnarray*}	
	\hat T_{j\to i} = \frac 1 {\det\vve C} \cdot 
		       \sum_{\nu=1}^d \Delta_{j\nu} C_{\nu,di}
			\cdot \frac {C_{ij}} {C_{ii}},
	\end{eqnarray*}
is replaced by (\ref{eq:Tji_est_new}),
	\begin{eqnarray*}
	\hat T_{j\to i} = 
	\frac 1 {\Delta t} \bracket{\log(\tilde {\vve C}^T \vve C^{-1})}_{ij}
		\cdot \frac {C_{ij}} {C_{ii}},
	\end{eqnarray*}
where $\vve {\tilde C} = (C_{i,j+})$, and $C_{i,j+}$ is the sample covariance 
between $x_i$ and $x_{j+}$, i.e., $x_j$ advanced by one time step.
Note here log is the matrix logarithm; in MATLAB, the function is logm.
This way, it shows that the preset
causality within the coupled system of chaotic oscillators has been rather
accurately reproduced even when the sampling interval is large (sampling
frequency is low).

	\begin{figure}[h]
	\begin{center}
	\includegraphics[width=0.95\textwidth]{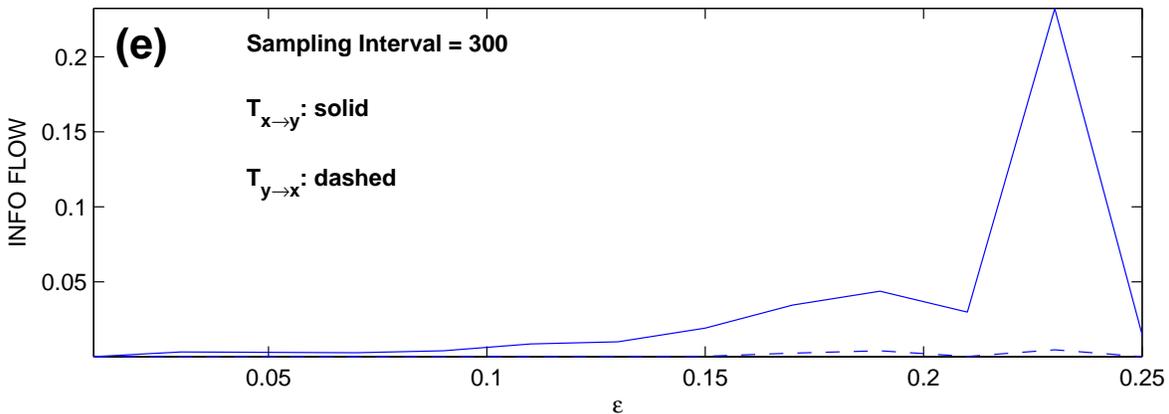}
	\caption{As Fig.~\ref{fig:new_IF_rossler}e, but 
	the covariances are estimated using the residuals of $x_i$ relative
	to the model result, instead of $x_i$ themselves. 
	Here SI=300 approximately corresponds to a sampling frequency 
	of 18 points each period.
	\protect{\label{fig:new_IF_SI300a}} }
	\end{center}
	\end{figure}	

There is still much room for improvement for the above approach.
For example, the estimation of the covariances in the quotient 
$\frac {\sigma_{ij}} {\sigma_{ii}}$ is by replacing the
population covariances with sample covariances, while the sample is formed
from the time series. While this is satisfactory
for stochastic systems under the ergodic assumption, this may not be good
for deterministic chaos, such as the R\"ossler oscillators case here.
The reason is obvious: 
The time mean of the series in Fig.~\ref{fig:xy_SI500} is zero, 
but one can imagine that the ensemble mean of all the possible paths 
is by no means zero; rather, it should be a function of time 
(just like the series itself), 
which may be close to the asymptotic linear system solution.
So it makes more sense to treat the linear system solution as the mean. 
As such, we have attempted to improve the estimation by replacing 
the covariances of $\ve x$ with those of $\ve x - \overline{\ve x}$, 
where $\overline{\ve x}$ stands 
for the resulting linear system solution;
With this we get another causal inference result 
for SI=300; the resulting IFs are plotted in Fig.~\ref{fig:new_IF_SI300a}. 
As one can see, the result looks rather accurate, just as expected, 
in contrast to Fig.~\ref{fig:new_IF_rossler}e. 
% This, and other potential methods for improvement, deserves for further investigation.

We, however, do not claim that we have solved the problem.
What we want to show here is, how much room there is for 
improvement within a linear framework. 
Indeed, in the case with high nonlinearity, the linear assumption 
is always easy to be blamed. 
While theoretically it is not a problem (causality is guaranteed as proved
in a theorem; see \cite{Liang2016} and other references), 
it is believed that, before a fully nonlinear algorithm is
developed, this will be a continuing issue.

\section*{Code availability}
The codes are available, and will be updated, at
www.ncoads.org/article/show/67.aspx.

\section*{Acknowledgments}
Many thanks for Milan Palu$\rm\check s$'s question, 
which motivated this research.

% \section*{Matlab codes}
% \begin{verbatim}
% \end{verbatim}

\end{document}